\pdfoutput=1
\documentclass[twocolumn,english,showpacs,preprintnumbers,amsmath,amssymb,pra]{revtex4}
\usepackage{amsmath}
\usepackage{graphicx}
\usepackage{amssymb}

\usepackage{color}

\begin{document}

\title{Cavity Quantum Electrodynamics with a Rydberg blocked atomic ensemble}

\author{Christine Guerlin$^{1,4}$, Etienne Brion$^{2}$, Tilman Esslinger$^{1}$
and Klaus M\o lmer$^{3}$}

\affiliation{$^1$Institute for Quantum Electronics, ETH Z\"urich,
CH-8093 Z\"urich, Switzerland, \\
$^2$Laboratoire Aim\'e Cotton, CNRS,
Campus d'Orsay, 91405, Orsay, France,\\ $^3$Lundbeck Foundation
Theoretical Center for Quantum System Research, Department of
Physics and Astronomy, University of Aarhus, DK-8000, Denmark,\\
$^4$Thales Research and Technology, Campus Polytechnique, 1 avenue
Augustin Fresnel, 91767 Palaiseau Cedex, France.}

\date{\today }
\begin{abstract}
We propose to implement the Jaynes-Cummings model by coupling a
few-micrometer large atomic ensemble to a quantized cavity mode and
classical laser fields. A two-photon transition resonantly couples
the single-atom ground state $\left|g\right\rangle $ to a Rydberg
state $\left|e\right\rangle $ via a non-resonant intermediate state
$\left|i\right\rangle $, but due to the interaction between Rydberg
atoms only a single atom can be resonantly excited in the ensemble.
This restricts the state space of the ensemble to the collective
ground state $\left|G\right\rangle $ and the collectively excited
state $\left|E\right\rangle $ with a single Rydberg excitation
distributed evenly on all atoms. The collectively enhanced coupling
of all atoms to the cavity field with coherent coupling strengths
which are much larger than the decay rates in the system leads to
the strong coupling regime of the resulting effective
Jaynes-Cummings model. We use numerical simulations to show that the
cavity transmission can be used to reveal detailed properties of the
Jaynes-Cummings ladder of excited states, and that the atomic
nonlinearity gives rise to highly non-trivial photon emission from
the cavity. Finally, we suggest that the absence of interactions
between remote Rydberg atoms may, due to a combinatorial effect,
induce a cavity-assisted excitation blockade whose range is larger
than the typical Rydberg dipole-dipole interaction length.
\end{abstract}


\maketitle

\section{Introduction}

The Jaynes-Cummings model (JCM) \cite{JC63} provides the general framework to describe the interaction of a two-level system -- such
as an atom, with a quantum harmonic oscillator -- e.g. a quantized cavity mode in cavity quantum electrodynamics (CQED). It allows
to explain very specific behaviors of such systems as, for example, the collapses and revivals of Rabi oscillations
\cite{KR82,Haroche}. The JCM was also used to describe various physical situations outside the field of CQED, such as the coupling
of the internal states to the centre-of-mass vibrational levels of a trapped ion subject to a laser beam in the Lamb-Dicke regime
(for a review, see \cite{LBMW03}), or, more recently, a laser-driven electron floating on liquid Helium \cite{ZJW09}. Precisely
because of this universality, the JCM also appears now as one of the key ingredients for applications in quantum information
processing.

While many successful CQED experiments have been carried out with
microwave cavities, resonant with transitions between atomic Rydberg
excited states \cite{haroche2006, Deleglise2008} and with
superconducting "artificial atoms" \cite{Wallraff2008,Schoelkopf},
it is harder to reach the strong coupling regime with single atoms
and optical cavities. This regime is, however, of considerable
practical interest, as it makes it possible for the cavity to work
as an interface between the optical photons, flying qubits, and the
atomic states, stationary qubits, in quantum computing and
communication networks \cite{Kimble}.

A viable path to obtain a highly coherent system in the optical regime is to use a collection of $N$ atoms whose coupling strength
to an optical field mode is magnified by the factor $\sqrt{N}$ relative to the single-atom case
\cite{nagorny2003,kruse2003,black2003,Klinner2006,slama2007,gupta2007,brennecke2007,colombe2007,Murch2008,brennecke2008,baumann2010}.
For large $N$, the collective excitation degree of freedom of an atomic ensemble is approximately equivalent to a quantum harmonic
oscillator and the system atoms-cavity therefore tends to be well-described by a simple quadratic Hamiltonian in the raising and
lowering operators for the atomic and field excitations. Concepts to squeeze and entangle the atomic and field degrees of freedom by
this Hamiltonian have been developed but the systems stay within the restricted family of Gaussian states
\cite{squeezing1,squeezing2}.

In this paper, we consider an atomic ensemble placed in an optical
high-finesse cavity and investigate the non-linearity introduced by
the interactions between atoms resonantly coupled to a high-lying
Rydberg state. In relatively small ensembles, dipole-dipole
interactions significantly shift the energy of states with two or
more Rydberg excitations \cite{walker2008}. When states with two
excited atoms are shifted far away from the optical resonance we
observe the so-called Rydberg Blockade phenomenon
\cite{Jaksch2000,Lukin2001,Tong2004,Singer2004,Saffman2005,Heidemann2007,
Vogt2006}, where the atomic ensemble effectively behaves as a single
two-level system. The entire atoms-cavity system is therefore
well-described by the Jaynes-Cummings model, and the collective
enhancement factor $\sqrt{N}$ compared to the single atom case may
then lead to the strong coupling regime.

The analyses we shall provide in this manuscript refer to a specific
experimental implementation with realistic physical parameters specified
in Sec. II. The expected signature of the strongly coupled ensemble
on the optical transmission of the cavity is discussed and investigated
numerically by Monte-Carlo simulations in Sec. III. In particular,
we show that both the average signal and the fluctuations present
interesting, and perhaps surprising, features linked with the eigenstates
of the Jaynes-Cummings model. In Sec. IV, we suggest a new mechanism
by which the cavity can extend the range of the Rydberg blockade beyond
the dipole-dipole interaction length. In Sec. V, we conclude and discuss
briefly the vast range of possible investigations that can be made
with the system and a few possible extensions of the theoretical model.

\section{Physical Implementation}

\begin{figure*}
\begin{centering}
\includegraphics[width=15cm]{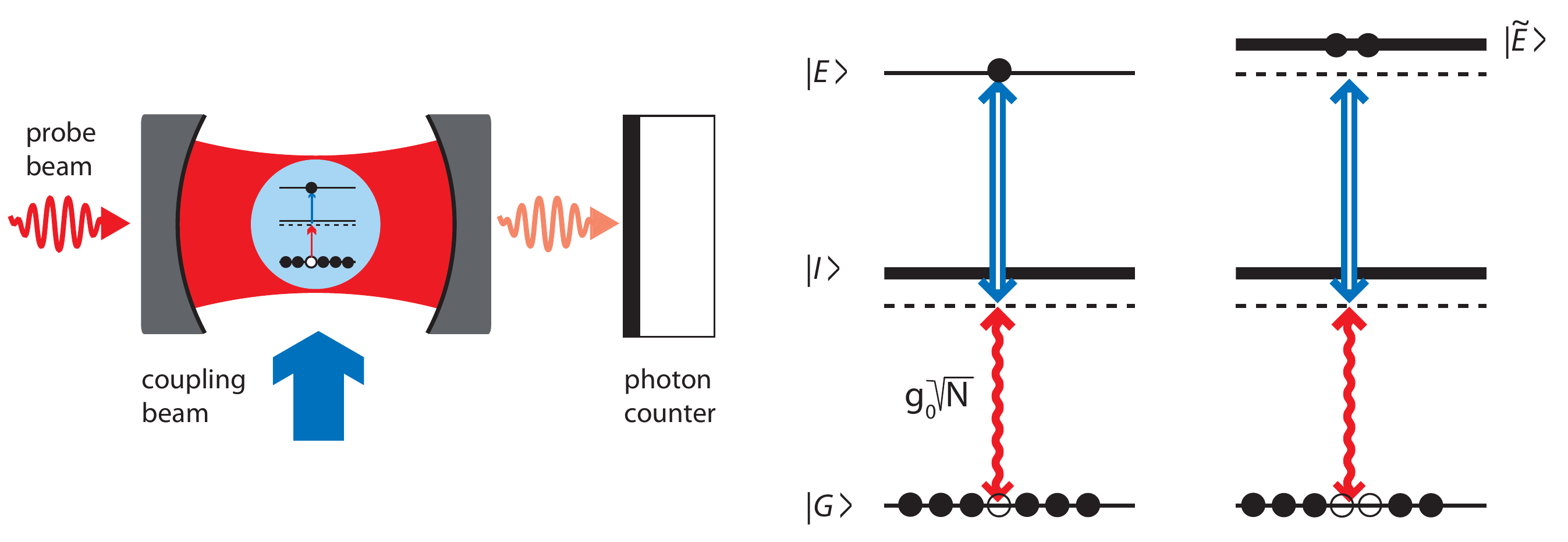}\\

\par\end{centering}

\caption{Left: A Bose-Einstein condensate inside an ultra-high finesse optical
cavity is coupled in a two photon scheme to a Rydberg state, with
the first photon being provided by a single quantized cavity mode
and the second photon by a classical coupling beam. The system is
interrogated by monitoring the cavity transmission of a weak probe
beam on a photon counter. Middle: The quantized cavity mode has a
frequency $\nu_{1}$ and couples the collective atomic ground state
$|G\rangle$ to an intermediate level $|I\rangle$ with a detuning
$\Delta$. The collective vacuum Rabi coupling is given by $g_{0}\sqrt{N}$.
A blue photon of frequency $\nu_{2}$ drives the excitation to a Rydberg
state $|E\rangle$ with a classical Rabi coupling $\Omega$. Right:
The interaction between two Rydberg atoms leads to an energy shifted
and broadened state $|\tilde{E}\rangle$ containing two excitations.
This allows selectively addressing the single excitation state. The
resulting situation can be described by a Jaynes-Cummings Hamiltonian.}

\label{Fig1}
\end{figure*}

The principle of our proposed experiment is depicted in Fig.
\ref{Fig1}. A Bose-Einstein condensate consisting of $N\gtrsim1000$
$^{87}$Rb atoms is placed in an ultra-high finesse cavity
\cite{brennecke2007} and is transversely illuminated by a
homogeneous classical laser field. The $5S_{1/2}$ ground atomic
state denoted by $\left|g\right\rangle $ is assumed to be resonantly
coupled to a Rydberg level $\left|e\right\rangle $ through a
two-photon process, via the $5P_{3/2}$ intermediate state
$\left|i\right\rangle $. The transition $\left|g\right\rangle
\rightarrow\left|i\right\rangle $, of frequency
$\omega_{g\rightarrow i}$, is non-resonantly coupled with the
coupling strength $g_{0}$ to a single quantized cavity mode with
annihilation operator $\hat{a}$ and frequency $\omega_{c}$. The
cavity mode is detuned by the amount
$\Delta=\omega_{c}-\omega_{g\rightarrow i}$ while the transition
$\left|i\right\rangle \rightarrow\left|e\right\rangle $ of frequency
$\omega_{i\rightarrow e}$, is non-resonantly driven by the laser
field with the Rabi frequency $\Omega$ and frequency $\omega_{l}$
detuned by the amount $\omega_{l}-\omega_{i\rightarrow e}=-\Delta$.
Omitting at first the interatomic interactions, one can describe the
physical situation by the following Hamiltonian \begin{eqnarray}
\tilde{H} & = & -\hbar\Delta\sum_{j=1}^{N}\left|i_{j}\right\rangle \left\langle i_{j}\right|+\left[\hbar g_{0}\hat{a}^{\dagger}\sum_{j=1}^{N}\left|g_{j}\right\rangle \left\langle i_{j}\right|+h.c.\right]\nonumber \\
 &  & +\left[\hbar\Omega\sum_{j=1}^{N}\left|e_{j}\right\rangle \left\langle i_{j}\right|+h.c.\right]\label{Ham1}\end{eqnarray}
 written in the interaction picture with respect to $H_{0}=\hbar\omega_{c}(\hat{a}^{\dagger}\hat{a}+\sum_{j=1}^{N}|i_{j}\rangle\langle i_{j}|)+\hbar(\omega_{c}+\omega_{l})\sum_{j=1}^{N}|e_{j}\rangle\langle e_{j}|$,
and in the rotating wave approximation. Dispersive shifts on both transitions have been neglected for simplicity. 
Assuming a large detuning from the intermediate state, $\Delta\gg\Gamma$,
where $\Gamma^{-1}$ is the lifetime of $\left|i\right\rangle $,
we can adiabatically eliminate the unpopulated intermediate state,
which leads to the effective Hamiltonian \begin{eqnarray}
\tilde{H}_{\mathrm{eff}} & = & \hbar\frac{g_{0}^{*}\Omega}{\Delta}\hat{a}\sum_{j=1}^{N}\left|e_{j}\right\rangle \left\langle g_{j}\right|+h.c.\nonumber \\
 & = & \hbar\sqrt{N}\frac{g_{0}^{*}\Omega}{\Delta}\hat{a}\hat{S}^{\dagger}+h.c.\label{Ham2}\end{eqnarray}
 where the collective mode atomic excitation is described by $\hat{S}^{\dagger}=\frac{1}{\sqrt{N}}\sum_{j=1}^{N}\left|e_{j}\right\rangle \left\langle g_{j}\right|$.

Eq. (1) is written under the assumption that all atoms have the same
coupling coefficients $g_0$, $ \Omega$ to the laser and cavity
field. This describes the situation of a Bose-Einstein condensate,
when neglecting the photon recoil. Alternatively, one can consider a
real space where the coupling strengths depends on the value of the
mode functions at the location of the individual atoms, and
$g^*_0\Omega$ should be replaced by different coefficients
$g^*_j\Omega_j$ on each term in the sum  in (2). It turns out,
however, that in the limit of many atoms, one can define a
collective operator like $\hat{S}^\dagger$ which is a weighted sum
of the individual atomic raising operators, and which also obeys the
oscillator like commutator relations to a good approximation. As
long as the atoms do not move appreciably and change their coupling
strengths on the time scale of interest for the experiment, this
weighted collective atomic mode plays the same role as the ideal
symmetric mode and its coupling is enhanced by the same factor
$\sqrt{N}$ and an extra mode dependent factor of order unity, see,
e.g., \cite{Wesenberg}.

For low Rydberg excitation numbers, $\left[\hat{S},\hat{S}^{\dagger}\right]\simeq1$,
and the Hamiltonian Eq. (\ref{Ham2}) approximately describes the
beam splitter coupling of two degenerate oscillators.

The dipole-dipole interactions we have omitted so far are strong only between neighbouring Rydberg atoms. Their effect on an atomic
sample whose size is at most of the order of the typical Rydberg-Rydberg dipole-dipole interaction length $\ell_{R}$ is to
considerably shift the energy of multiply Rydberg excited states. Driving the transition $\left|g\right\rangle
\rightarrow\left|e\right\rangle $ as described above, one can thus resonantly couple the collective ground state
$\left|G\right\rangle \equiv\left|g_{1},\ldots,g_{N}\right\rangle $ to the collective symmetric state with a single Rydberg
excitation $\left|E\right\rangle \equiv\hat{S}^{\dagger}\left|G\right\rangle $, only, as higher excited levels are too far detuned.
This constitutes the Rydberg blockade phenomenon \cite{Jaksch2000,Lukin2001}. Restricting the Hamiltonian Eq. (\ref{Ham2}) to the
physically relevant atomic subspace $\left\{ \left|G\right\rangle ,\left|E\right\rangle \right\} $ we get $\tilde{H'}_{eff}=\hbar g_{\mathrm{eff}}\hat{a}\left|E\right\rangle \left\langle G\right|+h.c.$ which corresponds to the interaction part of
the Jaynes-Cummings Hamiltonian describing the resonant interaction of the cavity field with a fictitious two-level {}``super atom''
\cite{vuletic2006}, with the effective coupling constant
$g_{\mathrm{eff}}\equiv\sqrt{N}g_{0}^{*}\Omega/\Delta$. We finally pass to a new rotating frame defined by the unitary state transformation $\exp\left\{-it\omega_{c}\left(\hat{a}^{\dagger}\hat{a} + \left|E\right\rangle \left\langle E\right|
\right)\right\}$
to obtain the full Jaynes-Cummings Hamiltonian
\begin{eqnarray}
\tilde{H}_{JC} & = & \hbar\omega_{c}\hat{a}^{\dagger}\hat{a}+\hbar\omega_{c}\left|E\right\rangle \left\langle E\right|\nonumber \\
 &  & +\left[\hbar g_{\mathrm{eff}}\hat{a}\left|E\right\rangle \left\langle G\right|+h.c.\right]\label{Ham4}\end{eqnarray}
 with an effectively equal excitation energy of the field and ensemble
atomic states.

The parameters of this Jaynes-Cummings model implementation can be tuned over a wide range. The collective coupling on the first
transition $\sqrt{N}g_0$ scales indeed as the square root of the atom number and can thus be varied in the experiment up to the GHz
range \cite{brennecke2007}. One possible scheme would be to excite the 70s Rydberg state from a 1000 atoms BEC, with a
cavity-intermediate state detuning of $\Delta=2\pi\times900$~MHz, a single atom resonant coupling on that transition of
$g_{0}=2\pi\times10$~MHz and a blue laser Rabi coupling $\Omega=2\pi\times30$~MHz. Excitation of the intermediate atomic state
$|i\rangle$ can be neglected here since the cavity detuning $\Delta$ is large compared to both the atomic radiative decay rate
$\Gamma=2\pi\times3$~MHz from the intermediate state and the one-photon collective coupling strength $\sqrt{N}g_{0}\sim
2\pi\times320$~MHz. The resulting two-photon coupling $g_{\mathrm{eff}}\sim2\pi\times10$~MHz is higher than both decay rates of the
cavity $\kappa=2\pi\times1.3$~MHz and of the Rydberg state $\gamma=2\pi\times0.55$~kHz. This system reaches the strong coupling
regime on the ground-to-Rydberg state two-photon transition. Within the BEC Thomas-Fermi radius $r_{TF}\sim 2~\mu\mbox{m}$
corresponding to an isotropic trapping frequency $2\pi\times 180$~Hz, the minimum Rydberg interaction shift over the atomic sample
is $\Delta_{RI}\sim2\pi\times200$~MHz, which spectrally separates the states with two or more Rydberg excited atoms from the singly
excited ones.

%

\section{The driven Jaynes-Cummings model}

In the previous section, we arrived at an effective description of
the ensemble-light interaction in terms of a two-level super atom
whose Hilbert space consists of the collective ground and singly
Rydberg excited states of the ensemble $\left|G\right\rangle
,\left|E\right\rangle $. The associated effective Jaynes-Cummings
Hamiltonian Eq. (\ref{Ham4}) is block diagonal, coupling only pairs
of states $\left(\left|G,n\right\rangle ,\left|E,n-1\right\rangle
\right)$ with the same total number $n$ of either photonic or atomic
excitations. The frequencies $\omega_{n,\pm}$ of the dressed
eigenstates $\left|n,\pm\right\rangle $, obtained by direct
diagonalization of $\tilde{H}_{JC}$, are given by

\begin{equation}
\omega_{n,\pm}=n\omega_{c}\pm g_{\mathrm{eff}}\sqrt{n}.\label{simple_JCspect}\end{equation}

\begin{figure*}
\begin{centering}
\includegraphics[width=15cm]{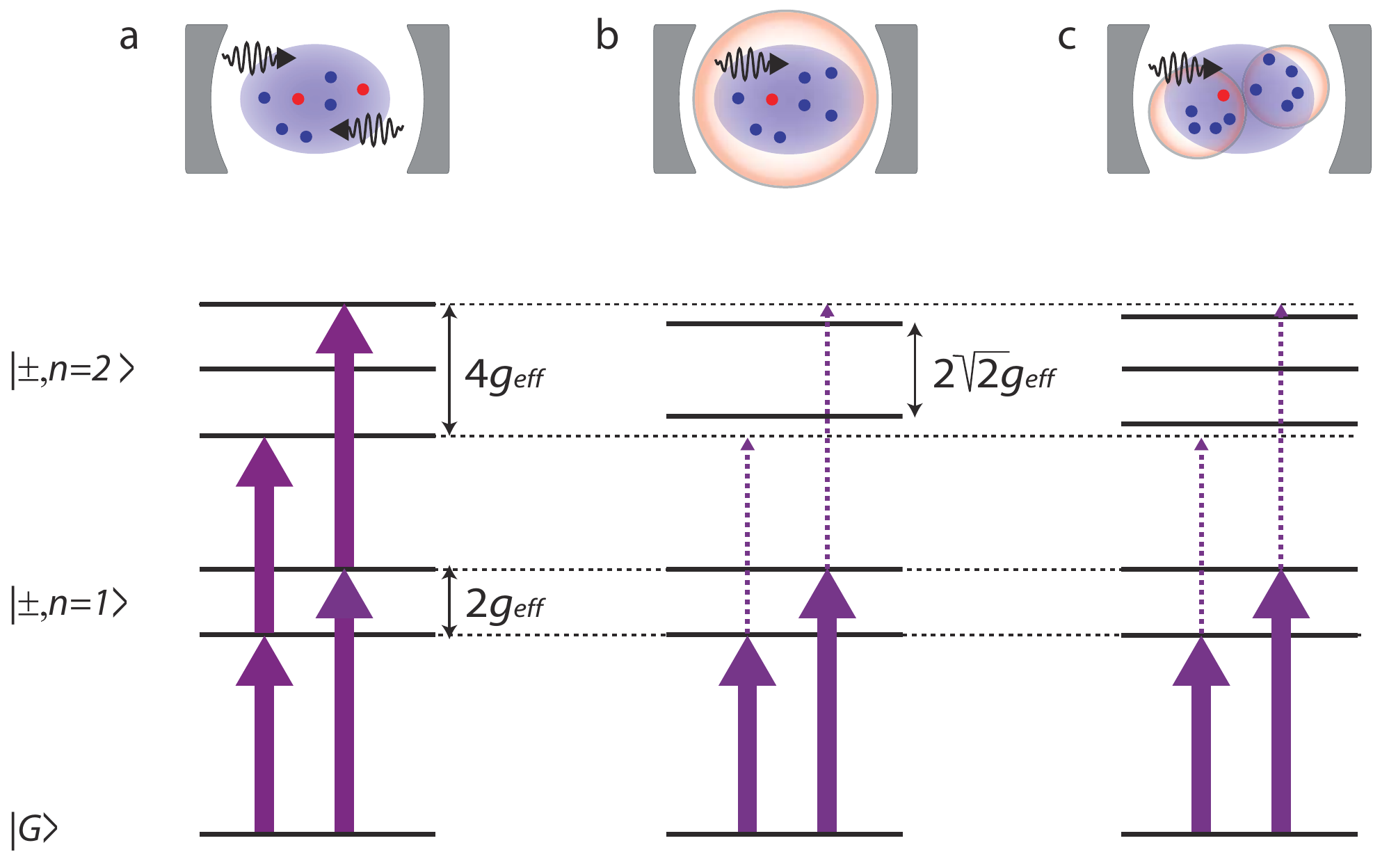}\\

\par\end{centering}

\caption{Dressed ladder states of the atoms-cavity system: a)
without inter-atomic interaction, the system would be formally
equivalent to two coupled oscillators and transition frequencies
between the multiplicities are degenerate; b) in case of interaction
between excited atoms strong enough for the atomic sample to be
fully dipole-blocked, the system is described by the JCM, and the
nonlinearity prevents simultaneous excitation of several
multiplicities; c) for larger atomic samples or weaker dipole
interaction, several Rydberg "bubbles" can appear, and the system's
spectrum is described by "combinatorially dressed states", still
preserving a nonlinearity for small enough bubble numbers.}

\label{Fig2}
\end{figure*}

\noindent The resulting nonlinearity is depicted on Fig. \ref{Fig2}.

%

This system can be driven by a classical probe field, which either
illuminates the atoms on the $g-i$ transition or feeds the cavity
via one of the cavity mirrors. In the following we shall consider
the latter solution. Due to the cavity and atomic decay, the
combined field-atom system will reach a steady state after a few
relaxation times, \emph{i.e.} a few microseconds for the specific
system under consideration. If the driving field is weak, this state
only slightly differs from the ground state
$\left|G,n=0\right\rangle $ and depending on the driving frequency,
we may estimate the probability amplitude with which the system
populates the excited eigenstates $\left|n,\pm\right\rangle $ of the
unperturbed Jaynes-Cummings model.

\subsection{One-photon resonance}

To excite the system, the driving frequency should match one of the
two dressed state components at $\omega_{c}\pm g_{\mathrm{eff}}$,
shifted by the so-called vacuum Rabi splitting. Directly measuring
the absorption or transmission of the cavity as a function of the
driving frequency
, thus reveals the collectively enhanced coupling strength
$g_{\mathrm{eff}}$. The width of the resonances is mainly governed
by the cavity decay (as noted above the atomic decay rate $\gamma$
is negligible compared to the cavity decay rate $\kappa$). With
$g_{\mathrm{eff}}\sim2\pi\times10\mbox{ MHz}$ and
$\kappa\sim2\pi\times1\mbox{ MHz}$, the two resonance lines are then
clearly split, and it is possible to selectively excite one of the
dressed states $\left|\pm,n=1\right\rangle
\equiv\left(\left|G,1\right\rangle \pm\left|E,0\right\rangle
\right)/\sqrt{2}$. If the probe field couples to the cavity mode
according to the Hamiltonian
\begin{align}
V=\alpha^{*}e^{i\omega t}\hat{a}+\alpha e^{-i\omega t}\hat{a}^{\dagger},\end{align}
 the coupling strength of the ground state to either of the dressed
states is the same, given by $\beta_{1}=\left\langle \pm,n=1\right|V\left|G,0\right\rangle =\alpha^{*}/\sqrt{2}$.
We therefore expect these states to be excited with a probability
$p_{1}\sim|\beta_{1}|^{2}/(\delta^{2}+\kappa^{2}/4)$, where $\delta$
denotes the probe frequency detuning with respect to the dressed state
eigenfrequency. The dressed state populated is a superposition of
the atomic and field excited state, and the mean photon number inside
the cavity is expected to be $\langle n\rangle=\frac{1}{2}p_{1}$.

To analyze the problem theoretically, we have carried out numerical
Monte Carlo simulations of the dynamics of the driven atom+cavity
system. The cavity decay occurs by emission of photons
\cite{Guerlin2007}, and we simulate the detection of these emission
events by a photon counter in the Monte Carlo Wave Function (MCWF)
formalism \cite{MCWF,Carmichael}. Such simulations do not converge
to a steady state, they rather present a stochastic dynamics with
click events followed by transient evolution until the next click
event is detected. When averaged over many independent realizations
of this stochastic process, one recovers the predictions of the
master equation \cite{Guerlin2007, Brune2008}. Moreover, the
individual stochastic simulations are also useful, as they provide
typical records of the randomly selected detection events which is,
indeed, information similar to the one obtained in a real
transmission experiment.

%
\begin{figure}
\begin{centering}
\includegraphics[width=8cm]{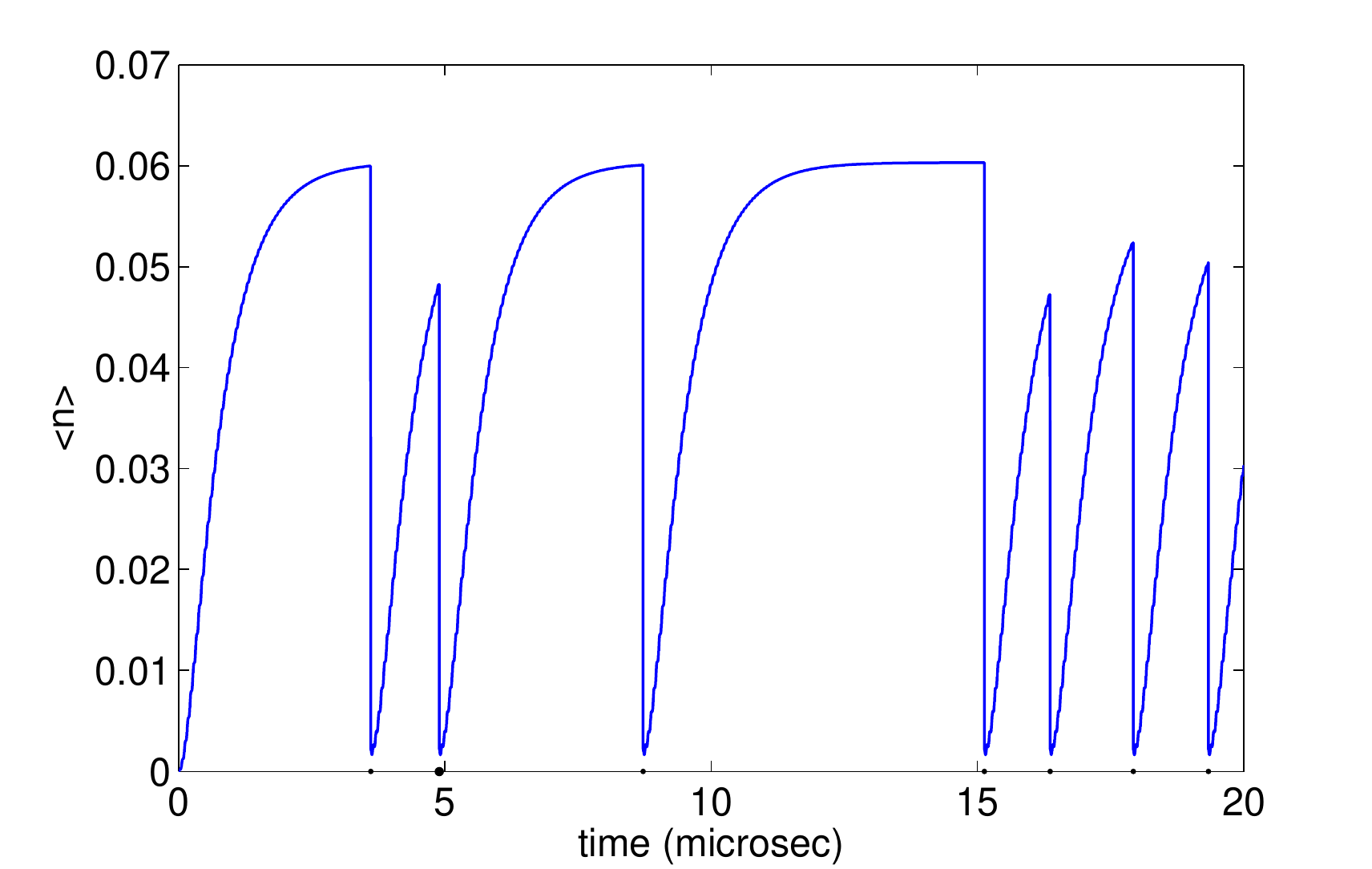}\\

\par\end{centering}

\caption{Detection record for one-photon resonance.
$\alpha=2\pi\times0.15\mbox{ MHz}$. }

\label{Fig3}
\end{figure}

Fig. \ref{Fig3} shows the result of a simulation, where we plot the
mean photon number in the cavity $\langle n\rangle$, as a function
of time. The graph shows characteristic oscillations at the Rabi
frequency of the transition, interrupted by sudden jumps
corresponding to the detection of a photon. At jumps the
intra-cavity photon number drops to nearly zero, while the state of
the system, in the dressed state superposition of $|G,1\rangle$ and
$|E,0\rangle$ just before the measurement, collapses to the ground
state $|G,0\rangle$. One also observes an immediate consequence of
these jumps: the system is unable to emit another photon until it
has been reexcited, hence the clicks on the detector are predicted
to arrive at a rate given by the mean excited state population and
to be antibunched on the time scale of $1/\beta_{1}$.

\subsection{Two-photon resonance}

Recently \cite{SKFPPMR08}, Schuster \emph{et al.} studied the
Jaynes-Cummings model of a single atom in an optical cavity, and
they identified a peak in the transmitted power, when the cavity is
driven at one of the frequencies $\omega_{c}\pm
g_{\mathrm{eff}}/\sqrt{2}$. Indeed, the driving field is then
resonant -- \emph{by two-photon absorption} -- with one of the
dressed states $\left|n=2,\pm\right\rangle $ with respective
energies $\omega_{2,\pm}=2\omega_{c}\pm g_{\mathrm{eff}}\sqrt{2}$.
This is a difficult experiment, both because this resonance is only
a few decay widths away from the single photon resonance, and
because the light atom interaction may cause heating of the atomic
motion in the frequency range giving the optimal resolution of the
resonance.

With our parameters, the splitting is larger, and due to the long
lifetime of the atomic Rydberg excited states, the light induced heating
will be much reduced. The second-order transition between the ground
state $|G,0\rangle$ and the dressed states $\left|\pm,n=2\right\rangle $
is detuned from the intermediate states $\left|\pm,n=1\right\rangle $
by the amount $\omega_{c}\pm g_{\mathrm{eff}}\sqrt{2}-\left(\omega_{c}\pm g_{\mathrm{eff}}/\sqrt{2}\right)=\pm g_{\mathrm{eff}}(1-1/\sqrt{2})$.
The two-photon coupling strength is hence estimated to be $\beta_{2}\sim\pm(\alpha/\sqrt{2})\alpha(1+\sqrt{2})/(g_{\mathrm{eff}}(1-1/\sqrt{2}))\sim\pm3\alpha^{2}/g_{\mathrm{eff}}$.
Accordingly, the steady state population of the doubly excited dressed
state is $p_{2}\sim|\beta_{2}|^{2}/(\delta^{2}+\kappa'^{2})$, where
$\kappa'$ denotes the coherence decay rate of the photonic components
of the dressed states. Since the dressed state is composed of equal
weight components with a single and two photons, the mean photon number
is expected to be $\langle n\rangle=\frac{3}{2}p_{2}$ and $\kappa'=3\kappa/2$
in this case.

%
\begin{figure}
\begin{centering}
\includegraphics[width=8cm]{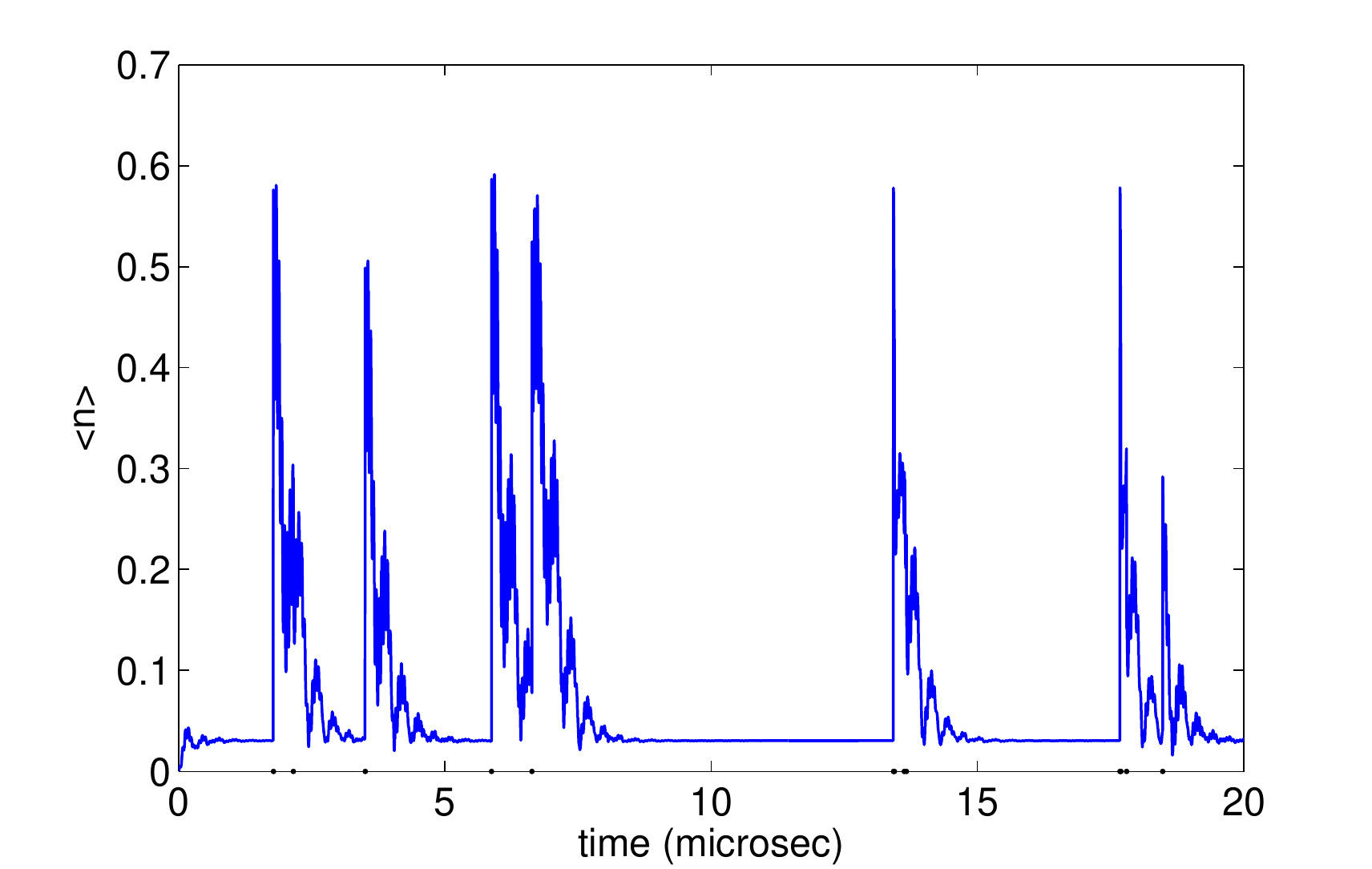}\\

\par\end{centering}

\caption{Detection record for two-photon resonance.
$\alpha=2\pi\times1.5\mbox{ MHz}$.}

\label{Fig4}
\end{figure}

Again, we carried out simulations which confirmed the existence of
the expected resonance. This time, however, the simulation record
looks very different from the results of the previous section, qs
shown on Fig. \ref{Fig4}. The mean photon number is low, and so is
the average transmitted flux, but every time a photon is detected,
we now see a drastic \emph{increase} in the mean photon number in
the cavity. This has a simple explanation, since the state prior to
the detection event is a superposition of the ground state of the
system and a doubly excited dressed state. The back action of the
detection of a single photon, the quantum jump, is implemented by
the action of the field annihilation operator $\hat{a}$, which
replaces the state before the detection event by \begin{eqnarray}
|\Psi\rangle_{jump}\propto\hat{a}\left\{ \left|G,0\right\rangle +\sqrt{p_{2}}(\left|G,2\right\rangle \pm\left|E,1\right\rangle )\right\} \nonumber \\
\propto\sqrt{2}\left|G,1\right\rangle \pm\left|E,0\right\rangle .\end{eqnarray}
 The mean photon number in this state is $2/3$, which is much larger
than the potentially nearly infinitesimal time averaged photon
number in the cavity. The figure shows the mean photon number
evaluated with the stochastic wave function method, and the
transient peaks confirm our simple analysis. The detection record
also shows a strong bunching effect: a majority fraction of the
jumps are followed by a second jump within the cavity lifetime. A
dedicated experiment should be able to verify these non-classical
intensity correlations, and in a future perspective, one may even
imagine the possibility to apply feed-back and modify the driving
field immediately after the first click event and thus prepare a
variety of other quantum states of the system \cite{Dotsenko2009}.

\subsection{Three-photon resonance}

More generally, the spectrum Eq. \ref{simple_JCspect} predicts the
existence of $n$-photon resonances at the driving frequencies $\omega_{n}=(n\omega_{c}\pm g_{\mathrm{eff}}\sqrt{n})/n=\omega_{c}\pm g_{\mathrm{eff}}/\sqrt{n}$.
For the specific experimental setup considered here, these resonances
are well distinguishable, even for $n>2$ : for example, the 3-photon
resonance is well separated by a few line widths from the 2- and 4-photon
resonances. Of course, the 3-photon excitation is a higher order process
with intermediate non-resonant 1- and 2-photon virtual excitations
of the system, and it competes with the non-resonant excitation of
the system by lower order processes. All these processes are included
in the full numerical simulations of the system.

%
\begin{figure}
\begin{centering}
\includegraphics[width=8cm]{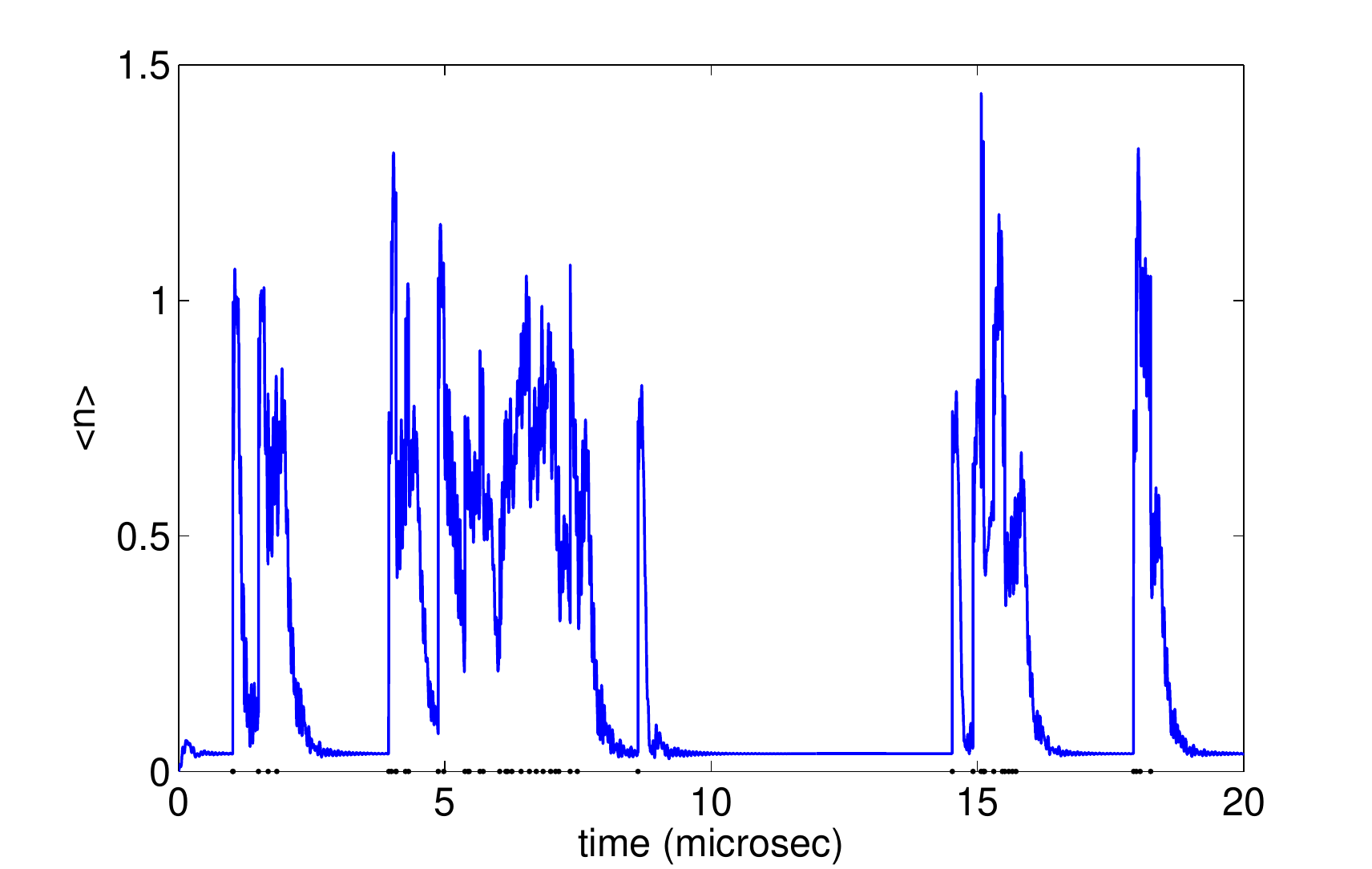}\\

\par\end{centering}

\caption{Detection record for three-photon resonance.
$\alpha=2\pi\times1.25\mbox{ MHz}$.}

\label{Fig5}
\end{figure}

Fig. \ref{Fig5} shows a simulated detection record for a driving
field resonant with the triply excited dressed state of the
atoms+cavity system. This time, our simple analysis suggests that
clicks cause a quantum jump of the state from a superposition of the
ground state and $(|G,3>\pm|E,2>)$ towards
$\sqrt{3}|G,2>\pm\sqrt{2}|E,1>$, which has an average photon number
of $8/5$. Within the cavity lifetime, one should thus expect bursts
of 2-3 photon detection events.

%
\begin{figure}
\begin{centering}
\includegraphics[width=8cm]{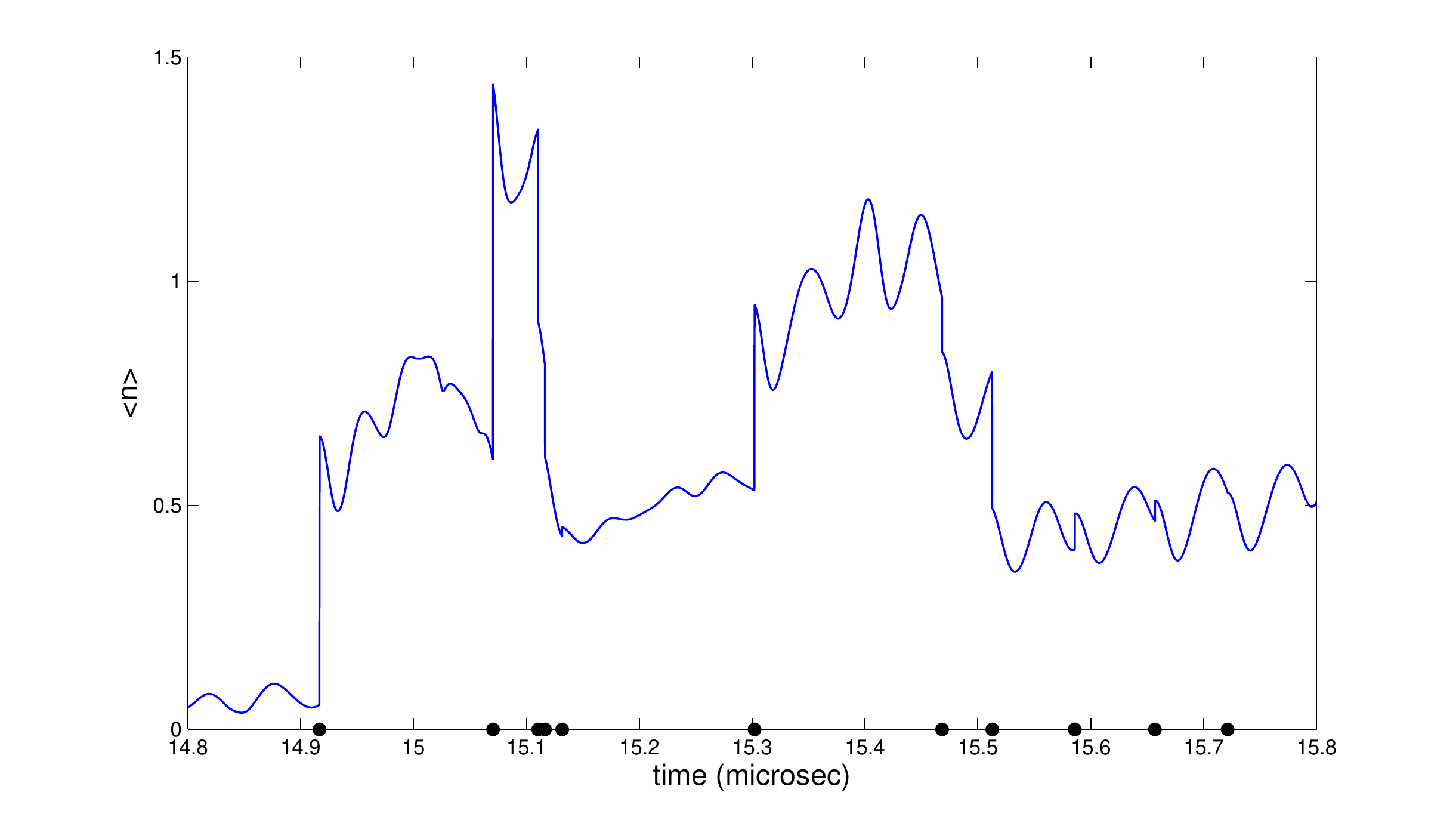}\\

\par\end{centering}

\caption{Detection record for three-photon resonance (magnified
view). Each click is indicated by a bullet symbol on the time axis.
$\alpha=2\pi\times2\mbox{ MHz}$.}

\label{Fig6}
\end{figure}

Such bunching of detection events is clearly seen in Fig.
\ref{Fig5}. But when we zoom in on the detection record (see Fig.
\ref{Fig6}), we observe somewhat surprising {}``bursts'' of 3, 4 and
even up to 7 detection events, which contradict our simple picture
of the dynamics. Looking more closely into the state conditioned on
the subsequent detection events, we have identified two mechanisms
which seem to contribute to the observed effect: \textit{i)} the
system does not only populate the ground and triply excited dressed
states, a small off-resonant excitation of higher lying states also
occurs. The corresponding population is then amplified by the action
of the annihilation operators and the $\sqrt{n}$ factors associated
with the detections of the first, second and third photon.
\textit{ii)} the system evolves between clicks, both due to the
driving field and due to the states of the form
$\sqrt{3}|G,2>\pm\sqrt{2}|E,1>$ resulting from the first click not
being eigenstates of the atoms-cavity coupling. Also, the small
probability amplitude in excited states, which is being magnified by
a quantum jump, contributes coherently to the build-up of further
excited state population by the coherent probe field. We have
qualitatively verified the last point by setting the probe amplitude
$\alpha$ to zero after the first click in our simulations and
observed a reduction of the number of transient detection events.

This multi-photon burst phenomenon constitutes an interesting experimental
effect, but we will not investigate it further at this point.

\section{Combinatorial extension of dipole blockade mechanism}

In the previous sections, we suggested to take advantage of the full
Rydberg blockade in a small atomic ensemble to effectively implement
the JCM in the strong coupling regime and we studied the transmission
properties of such an ensemble-cavity system. In this section, we
shall see how the coupling of the atoms to the cavity can actually
be used to extend the range of the blockade mechanism beyond the limit
fixed by the typical range $\ell_{R}\sim$ few $\mu m$ of Rydberg
dipole-dipole interaction.

In the absence of interactomic interactions, \emph{i.e.} for
ensembles with interatomic separations larger than $\ell_{R}$, the
atomic ensemble and the cavity mode behave like two coupled
oscillators leading to two atom-field eigenmodes with frequencies
$\omega_{c}\pm g_{\mathrm{eff}}$ (Fig. \ref{Fig2} a). Driving the
cavity with a classical probe field tuned on one of these
frequencies, one resonantly excites a coherent state of the
corresponding eigenmode equivalent to product of coherent states of
the field and the collective atomic oscillator comprising all the
Rydberg excitation number states. In the opposite limit, \emph{i.e.}
for ensembles whose linear dimension is of the order of $\ell_{R}$,
the dipole-dipole interactions between Rydberg atoms are so strong
that they shift the multiply Rydberg excited states out of
resonance: the ensemble is equivalent to a two-level atom, and the
entire system is described by the JCM, Eq.(\ref{Ham4}) (Fig.
\ref{Fig2} b).

Let us now turn to the intermediate regime where the extent of the
ensemble is larger than $\ell_{R}$ but the Rydberg dipole-dipole
interactions still does play an important role in the system. In that
case, the interaction does not shift all the multiply excited states
out of resonance and we may identify an unshifted doubly excited collective
state \begin{equation}
\left|E_{2}\right\rangle \equiv\frac{1}{\sqrt{\mathcal{A}}}\left(\sum_{|\vec{r}_{j}-\vec{r}_{k}|>\ell_{R}}\left|e_{j},e_{k}\right\rangle \left\langle g_{j},g_{k}\right|\right)\left|G\right\rangle \end{equation}
 in which the excited atoms $\left(j,k\right)$ are always too far
apart from each other to interact strongly. The normalization constant
$\mathcal{A}$ can be evaluated by counting the number of states $\left|e_{j},e_{k}\right\rangle $
which contribute to $\left|E_{2}\right\rangle $, \emph{i.e.} the
number of pairs of atoms in the ensemble which are separated by a
distance larger than $\ell_{R}$. This state is thus resonantly accessible
from the first excited state $\left|E\right\rangle $ via the two-photon
absorption from the cavity and the blue laser classical field. The
Hamiltonian Eq. (\ref{Ham4}) must therefore be complemented by the
term \begin{align*}
 & 2\hbar\omega_{c}\left|E_{2}\right\rangle \left\langle E_{2}\right|+\hbar g_{eff}\left\langle E_{2}\left|\hat{S}^{\dagger}\right|E\right\rangle \left|E_{2}\right\rangle \left\langle E\right|\hat{a}+h.c.\end{align*}
 More generally, due to the size of the sample, the resonant coupling
by the cavity mode and laser fields to higher multiply excited states
$\left|E_{3}\right\rangle ,\left|E_{4}\right\rangle ,\ldots$ in which
$n_{e}=3,4,\ldots$ Rydberg atoms are separated by more than $\ell_{R}$,
may be possible. We can visualize the system as if the Rydberg blockade
phenomenon decomposes the ensemble into a number of {}``bubbles''
of radius $\ell_{R}$ \cite{Bubble}. In each bubble only a single
Rydberg excitation is possible and the number of atoms which can be
simultaneously excited in the ensemble coincides with the number of
bubbles, $n_{b}\simeq\frac{V}{\frac{4}{3}\pi\ell_{R}^{3}}$. For a
mesoscopic sample which can thus accommodate $n_{b}$ Rydberg excitations,
the Hamiltonian Eq. (\ref{Ham4}) must therefore be complemented by
$\left(n_{b}-1\right)$ terms coupling $\left|E_{0}\right\rangle \equiv\left|G\right\rangle $
to $\left|E_{1}\right\rangle \equiv\left|E\right\rangle $, $\left|E_{1}\right\rangle $
to $\left|E_{2}\right\rangle $, $\left|E_{2}\right\rangle $ to $\left|E_{3}\right\rangle $,
$\ldots$ , $\left|E_{n_{b}-1}\right\rangle $ to $\left|E_{n_{b}}\right\rangle $.

The ensemble-cavity dynamics is thus governed by the Hamiltonian \begin{align}
H & =\hbar\omega_{c}\hat{a}^{\dagger}\hat{a}+\hbar\omega_{c}\left(\sum_{k=1}^{n_{b}}k\left|E_{k}\right\rangle \left\langle E_{k}\right|\right)\label{Hamk}\nonumber \\
 & +\hat{a}\left(\sum_{k=1}^{n_{b}}g_{k}\left|E_{k}\right\rangle \left\langle E_{k-1}\right|\right)+h.c.\end{align}

The coupling strengths $g_{k}$ can be easily evaluated in the
"bubble" picture, since each bubble provides an effective two-level
atom and the states $\left|E_{k}\right\rangle $ are the
symmetrically excited states of these effective two-level systems
with precisely $k$ excitations. These states, in turn, are
equivalent to effective angular momentum states $|j,m\rangle$ with
the quantum numbers $j=n_{b}/2$ and $m=-n_{b}/2+k$.

The coupling strengths $g_{k}$ are given by the well-known matrix
elements of the angular momentum raising operator multiplied by the
effective two-state coupling of each bubble, $\langle k|\hat{j}^{\dagger}|k-1\rangle$,
and we get \begin{eqnarray}
g_{k}=\sqrt{(2j-k+1)\cdot k}\cdot\sqrt{\frac{N}{n_{b}}}\frac{g_{0}^{*}\Omega}{\Delta}\nonumber \\
=\sqrt{\left(n_{b}-k+1\right)\cdot k}\sqrt{\frac{N}{n_{b}}}\frac{g_{0}^{*}\Omega}{\Delta}.\end{eqnarray}

Setting $k=1$, we recover the expected collective coupling $g_{eff}$
to the first excited states, while \begin{equation}
g_{2}=\sqrt{1-\frac{1}{n_{b}}}\sqrt{2}\sqrt{N}\frac{g_{0}^{*}\Omega}{\Delta},\end{equation}
 which vanishes for $n_{b}=1$ as it should, and which approaches
the expected $\sqrt{2}g_{eff}$ in the non-blocked oscillator limit
of large $n_{b}$.

Any finite value, and in practice a not too large value, of $n_{b}$
thus breaks the coupled oscillator picture, and hence the excitation
spectrum of the coupled system will not constitute an equidistant
ladder of states. In particular, the resonant excitation of the
first dressed states with a single Rydberg atom component is not
resonant with the next excitation step, and the system thus behaves
as if the Rydberg blockade extends over the entire system across
different "bubbles" (Fig. \ref{Fig2} c). We recall that the absence
of simultaneous excitation of several bubbles is not due to their
mutual interaction but due to the coupling strength to higher
excited states which is modified because of the number of pairs that
contribute to this excitation. The blockade behavior is thus due to
a combinatorial effect. Diagonalizing the Hamiltonian (\ref{Hamk}),
within the space of two excitations, we find the eigenvalues
$0,\pm\sqrt{2-\frac{1}{n_{b}}}\hbar\sqrt{2}g_{eff}$, and hence the
field resonant with the first dressed state $\left|+,1\right\rangle
$ is detuned from the two-photon resonance by
$2\left(1-\sqrt{1-\frac{1}{2n_{b}}}g_{eff}\right)\sim\frac{g_{eff}}{2n_{b}}$.

This analysis suggests that we can only extend the blockade over a
few, say less than ten, "bubbles". We intend to perform more
detailed numerical studies of the collectively excited atoms. The
picture with a given fixed set of bubbles of atoms is not an exact
one as any atom has the same amplitude to be excited and hence form
the center of a bubble of excluded excitations around it, and the
full many body state is a superposition of correspondingly different
bubble configurations \cite{bubble1,bubble2,bubble3}. We do expect,
however, that our combinatorial argument is robust to finer details
of the description and that any two-atom collective component will
not be excited at the same frequency as the dressed states with a
single collective excitation.

\section{Discussion}

We have suggested to use the collective coupling of $\sim1000$ atoms
to a cavity field combined with the Rydberg blockade effect to restrict
the atomic ensemble to an effective two-level system. Physical parameters
realistically reach the JCM of strong coupling CQED, and we have shown
that exploration of higher excited states of the dressed states ladder
of this model should be possible by optical transmission experiments.

In a longer perspective, many more possibilities can be
investigated. One may take advantage of the atomic sub-level
structure to significantly enrich the phenomenology with collective
qubit encoding \cite{Brion}, multi-mode atomic storage states,
STIRAP processes \cite{Moller}, and several quantum field components
may couple to Rydberg states with different pairwise interaction
properties \cite{Saffmancat} to accommodate effective optical
non-linearities \cite{Adams}.

In this work, we disregarded the role of atomic motion, but we note that even though the blockade regime precludes strong
interatomic forces as only single atoms are excited, the position-dependent collective field-atom coupling may induce mechanical
motion entangled with the field and atomic internal state degrees of freedom. If the atoms form a Bose-Einstein condensate of
interacting atoms, as in \cite{brennecke2007}, collective excitations will then act as micro-mechanical degrees of freedom
\cite{Murch2008,brennecke2008}, with potentially strong coupling to the collective internal state two-level system.

Another interesting direction of research, involves the examination
of larger systems. In such systems, the Rydberg blockade is only
efficient in blockade spheres around each atom, and remote atoms may
be simultaneously excited. We have proposed that the combinatorics
of the atoms-cavity coupling may distort the ladder of energies of
singly and multiply occupied Rydberg states, and thus maintain the
blockade effect over a larger atomic ensemble. This should be more
quantitatively addressed in future work.

\begin{acknowledgments}
We acknowledge stimulating discussions with Kristian Baumann,
Ferdinand Brennecke, Silvan Leinss and Simon Michels.

CG acknowledges support from an ETH fellowship. TE acknowledges
support by the the EU under FP7 FET-open (NameQuam) and the ERC
advanced grant SQMS. KM acknowledges support from IARPA through the
US Army Research Office Grant No. W911NF-05-1-0492.

\end{acknowledgments}

\bibliographystyle{apsrev}



\begin{thebibliography}{26}

\bibitem{JC63}E. T. Jaynes and F. W. Cummings, Proc. IEEE \textbf{51},
89 (1963).

\bibitem{KR82}P. L. Knight and P. M. Radmore, Phys. Rev. A \textbf{26},
676 (1982).

\bibitem{Haroche} M. Brune, F. Schmidt-Kaler, A. Maali, J. Dreyer, E. Hagley, J. M. Raimond, and S. Haroche, Phys. Rev. Lett. \textbf{76}, 1800 (1996).


\bibitem{LBMW03}D. Leibfried, R. Blatt, C. Monroe and D. Wineland,
Rev. Mod. Phys. \textbf{75}, 281 (2003).

\bibitem{ZJW09}M. Zhang, H. Y. Jia and L. F. Wei, Phys. Rev. A \textbf{80}, 055801 (2009).

\bibitem{haroche2006} S.~{H}aroche, J.~M. {R}aimond, \emph{{E}xploring the {Q}uantum}
(Oxford University Press, Oxford, 2006).

\bibitem{Deleglise2008} S. Del\'{e}glise, I. Dotsenko, C. Sayrin, J.
Bernu, M. Brune, J. M. Raimond, and S. Haroche, Nature \textbf{455},
510 (2008)

\bibitem{Wallraff2008} J. M. Fink, M. G\"{oš}ppl, M. Baur, R. Bianchetti, P. J. Leek, A. Blais, and A. Wallraff, Nature \textbf{454},
315(2008).

\bibitem{Schoelkopf} L. S. Bishop, J. M. Chow, J. Koch, A. A. Houck,
M. H. Devoret, E. Thuneberg, S. M. Girvin and R. J. Schoelkopf, Nature
Physics \textbf{5}, 105 - 109 (2009).

\bibitem{Kimble} H. J. Kimble, Nature \textbf{453}, 1023-1030, (2008).

\bibitem{nagorny2003}
B.~{N}agorny, {T}h. {E}ls\"asser, A.~{H}emmerich, {P}hys. {R}ev.
{L}ett. \textbf{91}, 153003 (2003).

\bibitem{kruse2003}
D.~{K}ruse, C.~von {C}ube, C.~{Z}immermann, P.~W. {C}ourteille,
{P}hys. {R}ev. {L}ett. \textbf{91}, 183601 (2003).

\bibitem{black2003}
A.~T. {B}lack, H.~W. {C}han, V.~{V}uleti\'{c}, {P}hys. {R}ev.
{L}ett. \textbf{91}, 203001 (2003).

\bibitem{Klinner2006}
J. Klinner, M. Lindholdt, B. Nagorny, and A. Hemmerich, {P}hys.
{R}ev. {L}ett. \textbf{96}, 23002 (2006).

\bibitem{slama2007}
S.~{S}lama, S.~{B}ux, G.~{K}renz, C.~{Z}immermann, P.~W. {C}ourteille,
  {P}hys. {R}ev. {L}ett.  {\bf 98}, 053603 (2007).

\bibitem{gupta2007}
S.~{G}upta, K.~L. {M}oore, K.~W. {M}urch, D.~M. {S}tamper-{K}urn, {P}hys. {R}ev. {L}ett. \textbf{99}, 213601 (2007).

\bibitem{brennecke2007}
F.~{B}rennecke, T.~{D}onner, S.~{R}itter, T.~{B}ourdel, M.~{K}\"ohl,
  T.~{E}sslinger, {N}ature \textbf{450}, 268 (2007).

\bibitem{colombe2007}
Y.~{C}olombe, T.~{S}teinmetz, G.~{D}ubois, F.~{L}inke, D.~{H}unger,
  J.~{R}eichel, {N}ature \textbf{450}, 272 (2007).

\bibitem{Murch2008}
K. W. Murch, K. L. Moore, S. Gupta, D. M. Stamper Kurn, Nat. Phys.
\textbf{4}, 561 (2008).

\bibitem{brennecke2008} F.~Brennecke, S.~Ritter, T.~Donner, T.~Esslinger,
Science \textbf{322}, 235 (2008)

\bibitem{baumann2010} K. Baumann, C. Guerlin, F. Brennecke, T. Esslinger, Nature \textbf{464}, 1301 (2010).

\bibitem{squeezing1} A. S\o rensen and K. M\o lmer, Phys. Rev.
A. \textbf{66}, 022314 (2002); L. B. Madsen and K. M\o lmer, Phys. Rev. A \textbf{70}, 052324 (2004).

\bibitem{squeezing2}M. H. Schleier-Smith, I. D. Leroux, and
V. Vuleti\'{c}, Phys. Rev. A \textbf{81}, 021804 (2010); I. D. Leroux, M. H. Schleier-Smith, and V. Vuleti\'{c}, Phys. Rev. Lett.
\textbf{104}, 073602 (2010).

\bibitem{walker2008} T. G.~{W}alker, M.~{S}affman, {P}hys. {R}ev. A \textbf{77}, 032723
(2008).

\bibitem{Jaksch2000} D. Jaksch, J. I. Cirac, P. Zoller, S. L. Rolston, R. C\^{o}t\'{e}, and M. D. Lukin, Phys. Rev. Lett. \textbf{85}, 2208 (2000).

\bibitem{Lukin2001} M. D. Lukin, M. Fleischhauer, R. C\^{o}t\'{e}, L. M. Duan, D. Jaksch, J. I. Cirac, and P. Zoller, Phys. Rev. Lett. \textbf{87}, 037901 (2001).

 \bibitem{Tong2004} D. Tong, S. M. Farooqi, J. Stanojevic, S. Krishnan, Y. P. Zhang, R. C\^{o}t\'{e}, E. E. Eyler, and P. L. Gould, Phys. Rev. Lett \textbf{93}, 063001 (2004).

 \bibitem{Singer2004} K. Singer, M. Reetz-Lamour, T. Amthor, L. G. Marcassa, and M. Weidem{\"u}ller, Phys. Rev. Lett. \textbf{93}, 163001 (2004).

 \bibitem{Saffman2005} M. Saffman and T. G. Walker, Phys. Rev. A \textbf{72}, 022347 (2005).

 \bibitem{Heidemann2007} R. Heidemann, U. Raitzsch, V. Bendkowsky, B. Butscher, R. L{\"o}w,
L. Santos, and T. Pfau, Phys. Rev. Lett. \textbf{99}, 163601 (2007).

\bibitem{Vogt2006} T. Vogt, M. Viteau, J. Zhao, A. Chotia, D. Comparat, and P. Pillet,
Phys. Rev. Lett. \textbf{97}, 083003 (2006).

\bibitem{vuletic2006} V.~Vuletic, Nat. Phys. \textbf{2}, 801 (2006).

\bibitem{Wesenberg} J. H. Wesenberg, A. Ardavan, G. A. D. Briggs, J. J. L. Morton, R. J. Schoelkopf, D. I. Schuster, K. M\o lmer, Phys. Rev. Lett. \textbf{103}, 070502 (2009).

\bibitem{Guerlin2007} C. Guerlin, J. Bernu, S. Del\'{e}glise, C. Sayrin,
S. Gleyzes, S. Kuhr, M. Brune, J. M. Raimond, and S. Haroche, Nature
\textbf{448}, 23 (2007).

\bibitem{MCWF} J. Dalibard, Y. Castin, and K. M\o lmer, Phys. Rev. Lett. \textbf{68}, 580 (1992); K. M\o lmer, Y. Castin, and J. Dalibard, J. Opt. Soc. Am B. \textbf{10}, 524 (1993).

\bibitem{Carmichael} H. Carmichael, \textit{An Open Systems Approach to Quantum Optics}, Lecture Notes in Physics, Vol \textbf{18} (Springer, Berlin, 1993).

\bibitem{Brune2008} M. Brune, J. Bernu, C. Guerlin, S. Del\'{e}glise, C.
Sayrin, S. Gleyzes, S. Kuhr, I. Dotsenko, J. M. Raimond, and S.
Haroche, Phys. Rev. Lett. \textbf{101}, 240402 (2008).

\bibitem{SKFPPMR08}
I. Schuster, A. Kubanek, A. Fuhrmanek, T. Puppe,
P. W. H. Pinkse, K. Murr and G. Rempe, Nat. Phys. \textbf{4}, 382
(2008).

\bibitem{Dotsenko2009}
I. Dotsenko, M. Mirrahini, M. Brune, S. Haroche, J.-M. Raimond, and
P. Rouchon, Phys. Rev. A \textbf{80}, 013805 (2009)

\bibitem{Bubble} T. Cubel Liebisch, A. Reinhard, P. R. Berman, and
G. Raithel, Phys. Rev. Lett. \textbf{95}, 253002 (2005).

\bibitem{bubble1} J. V. Hernandez and F. Robicheaux,  J. Phys. B \textbf{39}, 4883 (2006).

\bibitem{bubble2} H. Weimer, R. L\"{o}w, T. Pfau, and H. P. B\"{u}chler,
Phys. Rev. Lett. \textbf{101}, 250601 (2008).

\bibitem{bubble3} G. Pupillo, A. Micheli, M. Boninsegni, I. Lesanovsky, P. Zoller, \emph{Strongly correlated gases of Rydberg-dressed atoms: quantum and classical dynamics}; arXiv:1001.0519v1.

\bibitem{Brion} E. Brion, K. M\o lmer, M. Saffman, Phys. Rev. Lett.
\textbf{99}, 260501 (2007).

\bibitem{Moller} D. M\o ller, L. B. Madsen, and K. M\o lmer, Phys.
Rev. Lett. \textbf{100}, 170504 (2008).

\bibitem{Saffmancat} M. Saffman and K. M\o lmer, Phys. Rev. Lett.
\textbf{102}, 240502 (2009).

\bibitem{Adams} J. D. Pritchard, A. Gauguet, K. J.Weatherill, M. P.
A. Jones, and C. S. Adams, \textit{Cooperative optical nonlinearity
due to dipolar interactions in an ultra-cold Rydberg ensemble,} arXiv:0911.3523,
(2009).

\end{thebibliography}

\end{document}